\RequirePackage{fix-cm}
\documentclass[twocolumn,epjc3]{svjour3}  
\RequirePackage[T1]{fontenc}
\smartqed  
\RequirePackage{graphicx}
\RequirePackage{mathptmx}
\RequirePackage{latexsym}
\RequirePackage[dvipdfm,colorlinks,citecolor=blue,urlcolor=blue,linkcolor=blue]{hyperref}

\usepackage{cite}
\usepackage{atlasphysics}
\usepackage{subfigure}
\usepackage{array}
\usepackage{xspace}

\def\mzero{m_{0}}
\def\mth{m_{3/2}}
\def\nTRT{N_\mathrm{{TRT}}^{\mathrm{outer}}}
\def\ptem{\ensuremath{p_{\mathrm{T}}^{\mathrm{jet,EM}}}}

\journalname{Eur. Phys. J. C}

\begin{document}

\title{
  {\normalsize{CERN-PH-EP-2011-210}} \hspace{9.7cm}{\normalsize\normalfont{Submitted to Eur. Phys. J. C}}
  \\ \\
  {Search for anomaly-mediated supersymmetry breaking with the ATLAS detector based on a disappearing-track signature in $pp$ collisions at $\rts = 7\TeV$}
}

\author{The ATLAS Collaboration\thanksref{email,addr}}
\thankstext{email}{e-mail: atlas.publications@cern.ch}
\institute{
CERN, 1211 Geneva 23, Switzerland\label{addr}
}
\date{Received: date / Accepted: date}
%
\maketitle

\begin{abstract}
In models of anomaly-mediated supersymmetry breaking (AMSB), the lightest chargino is predicted to have a lifetime long enough to be detected in collider experiments.
This letter explores AMSB scenarios in $pp$ collisions at $\rts=7\TeV$ by attempting to identify decaying charginos which result in tracks that appear to have few associated hits in the outer region of the tracking system.
The search was based on data corresponding to an integrated luminosity of $1.02~\ifb$ collected with the ATLAS detector in 2011.
The $\pt$ spectrum of candidate tracks is found to be consistent with the expectation from Standard Model background processes and constraints on the lifetime and the production cross section were obtained.
In the minimal AMSB framework with $m_{3/2}<32\TeV$, $m_0<1.5\TeV$, $\tan\beta=5$ and $\mu>0$, a chargino having mass below $92\GeV$ and a lifetime between 0.5 ns and 2 ns is excluded at 95\% confidence level.
\end{abstract}

\section{Introduction}
\label{sec:introduction}
Supersymmetry (SUSY)~\cite{Golfand:1971iw,Neveu:1971rx,Neveu:1971iv,Ramond:1971gb,Volkov:1973ix,Wess:1973kz,Wess:1974tw,Fayet:1977yc,Farrar:1978xj} is a promising solution to the hierarchy problem of the Standard Model (SM) and the search for SUSY is an important programme at the Large Hadron Collider (LHC).
For each SM particle, SUSY postulates a supersymmetric partner with identical quantum numbers but with a spin that differs by 1/2.
Since scalar superpartners of quarks and leptons with masses equal to quarks and leptons have not been observed in previous searches, SUSY must be a broken symmetry.
One mechanism which provides a calculable mass spectrum of supersymmetric particles is provided by anomaly mediation~\cite{Randall:1998uk,Giudice:1998xp}.
The anomaly-mediated SUSY breaking (AMSB) model provides a constrained particle mass spectrum; the ratios of the three gaugino masses are given approximately as 
$M_1 : M_2 : M_3 \approx 3 : 1 : 7$ where $M_i$ $(i=1,2,3)$ are the bino, wino and gluino masses, respectively.
The neutral wino becomes the lightest supersymmetric particle (LSP) while the charged wino becomes slightly heavier due to radiative corrections involving electroweak gauge bosons in the loops.
This phenomenological feature of the nearly degenerate lightest chargino ($\chinoonepm$) and neutralino (\ninoone) has the important implication that
$\chinoonepm$ predominantly decays into $\ninoone$ plus a low-momentum ($\sim$ 100 $ \MeV$) $\pi^{\pm}$.
The decay length of $\chinoonepm$ is typically expected to be a few centimeters at LHC energies; some $\chinoonepm$ charginos could therefore decay inside the tracking volume of the ATLAS detector.
The $\ninoone$ escapes detection and the softly emitted $\pi^{\pm}$ is not reconstructed.
A track arising from a $\chinoonepm$ with these characteristics is classified as a disappearing track.
The search described in this letter is based on this signature of decaying charginos which leads to a track having few associated hits in the outer part of the tracking volume.

\section{The ATLAS detector}

ATLAS is a multi-purpose detector~\cite{Aad:2008zzm}, covering nearly the entire solid angle\footnote{
ATLAS uses a right-handed coordinate system with its origin at the nominal interaction point (IP) in the centre of the detector and the $z$-axis coinciding with the axis of the beam pipe.
The $x$-axis points from the IP to the centre of the LHC ring, and the $y$-axis points upward.
Cylindrical coordinates ($r,\phi$) are used in the transverse plane, $\phi$ being the azimuthal angle around the beam pipe.
The pseudorapidity is defined in terms of the polar angle $\theta$ as $\eta = -\ln \tan(\theta/2)$.
}
around the collision point with layers of inner tracking devices surrounded by a superconducting solenoid providing a $2~\mathrm{T}$ magnetic field, a calorimeter system and a muon spectrometer.
The inner tracking detector provides tracking in the region $|\eta|<2.5$.
It consists of pixel and silicon microstrip (SCT) detectors inside a transition radiation tracker (TRT).

Of particular importance to this analysis is the TRT which covers the region $|\eta|<2.0$.
The barrel TRT is divided into inner, middle and outer concentric rings of 32 modules comprising a stack in the azimuthal angle; each covers the radial range from $563~\mathrm{mm}$ to $1066~\mathrm{mm}$ and $|\eta|<1.0$.
A module consists of a carbon-fibre laminate shell and an array of straw tubes and has a different structure for each ring.

The calorimeter system covers the range $|\eta|<4.9$.
The electromagnetic calorimeter is a lead/liquid-argon (LAr) detector in the barrel ($|\eta| < 1.475$) and endcap ($1.375 < |\eta| < 3.2$) regions.
The hadron calorimeters are composed of a steel and scintillator barrel ($|\eta|<1.7$), a LAr/copper endcap ($1.5<|\eta|<3.2$) and a LAr forward system ($3.1<|\eta|<4.9$) with copper and tungsten absorbers.
The muon spectrometer consists of three large superconducting toroids with 24 coils, a system of trigger chambers and precision tracking chambers which provide muon momentum measurements up to $|\eta|$ of 2.7.

\section{Simulated event samples}

Simulated Monte Carlo (MC) events were used to assess the experimental sensitivity to given models.
The minimal AMSB model is characterized by four parameters: the gravitino mass ($m_{3/2}$), the universal scalar mass ($m_0$), the ratio of Higgs vacuum expectation values at the electroweak scale ($\tan\beta$) and the sign of the higgsino mass term ($sgn(\mu)$). 
In this letter, {\sc Isasusy} from {\sc Isajet} v7.80~\cite{Paige:2003mg} was used to calculate the SUSY mass spectrum and the decay tables.
The MC samples were produced using {\sc HERWIG++}~\cite{Herwig++} with {\sc MRST2007 LO*}~\cite{MRST2007} parton distribution functions.
These samples were produced using the parameter tune described in Ref. \cite{ATLAS-CONF-2010-031} and a detector simulation based on {\sc Geant4} \cite{Agostinelli:2002hh,Aad:2010wqa} with multiple $pp$ interactions per event (pile-up) to match what was observed in data.
Given the chargino mass ($m_{\chinoonepm}$) limit by the LEP2 searches~\cite{Heister:2002mn,OPAL:AMSB,Alderweireld:999530} of $m_{\chinoonepm} \simeq 92\GeV$ at 95\% confidence level (CL), the signal models shown in Table~\ref{table:signal_parameters} were tested.
A large value of $m_0$ was used in order to prevent the existence of a tachyonic slepton;
this also assigns heavy masses to the squarks and sleptons, thereby avoiding constraints from flavour-changing neutral current and CP-violation measurements.
In this search, the production processes $\tilde{g}\tilde{g}$, $\tilde{q}\tilde{g}$ and $\tilde{q}\tilde{q}$ were considered.
The signal samples were normalized using next-to-leading-order (NLO) cross sections determined with {\sc Prospino}~\cite{Beenakker:1996ch}.
The chargino lifetime ($\tau_{\chinoonepm}$) was set to $1~\mathrm{ns}$, the value for which this analysis has the highest sensitivity.
The branching fraction for the decay $\chinoonepm \to \ninoone \pi^{\pm}$ was set to 100\%.
Samples with different lifetime values for each signal model were derived by applying event weights so that the distribution of the proper lifetime follows that for a given lifetime value.

In the model, gluinos and squarks are expected to be produced copiously via the strong interaction in $pp$ collisions.
The decay cascade of these to the $\chinoonepm$ and $\ninoone$ produces multiple jets with high transverse momentum ($\pt$).
LSPs escape from the detector, resulting in an event topology with multiple jets and large missing transverse momentum ($\met$).
Chargino tracks are expected to have significant transverse momentum since the difference between the gluino and chargino masses is large;
the chargino track typically has $\pt>50\GeV$ and is well isolated from the jet activity in the event. 

\begin{table}
\caption{Summary of AMSB signal parameters, chargino masses and their NLO cross sections with $\tan\beta=5$ and $sgn(\mu)=+1$.}
\label{table:signal_parameters}
\begin{tabular*}{\columnwidth}{@{\extracolsep{\fill}}ccccc@{}}
\hline
\hline
Signal & $\mzero$[$\mathrm{TeV}$] & $\mth$[$\mathrm{TeV}$] & $m_{\chinoonepm}$[$\mathrm{GeV}$] & Cross section[$\mathrm{pb}$] \\
\hline
LL01 & 1.5 & 32 &  90.2 & $6.10 \times 10^{-2}$ \\
LL02 & 1.8 & 41 &  117.8 & $7.65 \times 10^{-3}$  \\
LL03 & 2.0 & 51 &  147.7 & $1.00 \times 10^{-3}$  \\
\hline
\hline
\end{tabular*}
\end{table}

\section{Data and event selection}
The analysis was based on $pp$ collision data at $\rts=7\TeV$ recorded from March to July 2011.
The corresponding integrated luminosity, after the application of beam, detector and data quality requirements, was $1.02~\ifb$.
Events were selected at the trigger level by requiring at least one jet with a transverse momentum, measured at the electromagnetic scale, above $75\GeV$, and a missing transverse momentum above $55\GeV$.

Jets were reconstructed using the anti-$k_{t}$ algorithm~\cite{Cacciari:2008gp} with a distance parameter of 0.4.
The inputs to the jet reconstruction algorithm were three-dimensional topological calorimeter energy clusters~\cite{Lampl:1099735}.
The measurement of jet transverse momentum at the electromagnetic scale ($\ptem$) underestimates the true momentum due to the nature of the non-compensating calorimeters and the dead material.
Thus, an average correction~\cite{ATLAS-JET-ENERGY}, depending on $\eta$ and $\ptem$, was applied to obtain the calibrated jet $\pt$.
Jets with $\pt>20\GeV$ and $|\eta|<3.2$ were selected.
Electron candidates were selected with "medium" purity cuts, as described in Ref.~\cite{Aad:2010yt}.
Furthermore, electrons were required to fulfill the requirements of $\pt>10\GeV$, $|\eta|<2.47$ and $\sum_{\Delta R<0.2}\pt^{\mathrm{track}}/\pt$ $<$ $0.1$, where $\sum_{\Delta R<0.2}\pt^\mathrm{track}$ is the sum of $\pt$ for all the tracks with $\pt>1\GeV$ in a cone of $\Delta R$ $\equiv$ $\sqrt{\left( \Delta\eta \right)^2 + \left( \Delta\phi \right)^2}$ $< 0.2$ around the electron candidate, excluding the $\pt$ of the electron candidate itself.
Muon candidates were identified by an algorithm which combines a track reconstructed in the muon spectrometer with a track in the inner detector.
Furthermore, muons were required to have $\pt>10\GeV$ and $|\eta|<2.7$, and to be isolated \cite{ATLAS:2011ad}: the sum of $\pt$ of tracks within a cone of $\Delta R$ $< 0.2$ around the muon candidate (excluding the muon candidate itself) was required to be less than $1.8 \GeV$.

Following the object reconstruction described above, overlaps between jets and leptons were resolved.
First, any jet candidate lying within a distance of $\Delta R$ $< 0.2$ of an electron was discarded.
Then, any lepton candidates within a distance of $\Delta R<0.4$ of any surviving jet were discarded.

The calculation of $\met$ was based on the transverse momenta of jets and lepton candidates, and all clusters in the calorimeter that are not associated to such objects~\cite{Aad:2012re}.

In order to suppress non-collision background events, additional selection criteria~\cite{ATLAS-JET-ENERGY} were applied to jets.
Signal candidate events were required to have no electron or muon candidates (lepton veto), $\met>130\GeV$ and three leading (highest $\pt$) jets with $\pt>130\GeV$ for one jet and $\pt>60\GeV$ for another two jets ("kinematic selection").
The trigger selection is fully efficient for signal events satisfying the kinematic selection requirements.

The search described in this letter was based on the detection of charginos decaying in the TRT.
The average number of hits on a track going through the TRT in the central region is about 34 and consecutive hits can be observed along the track with small radial spacing between adjacent hits.
This feature provides the capability of substantial discrimination between penetrating and decaying charged particles.
If a chargino decays in the volume of the inner or middle TRT modules, multiple hits associated to the chargino track are expected in the SCT detector but not in the outer TRT subdetector.
Such a chargino track candidate can be fully reconstructed by the ATLAS standard track reconstruction algorithm.

The chargino candidate tracks were required to fulfill the following criteria:
\begin{itemize}
  \item[(1)] The track should have at least one hit in the innermost layer of the pixel detector.
  \item[(2)] The track should have at least six hits in the SCT.
  \item[(3)]  The track should have $|d_0| < 1.5~\mathrm{mm}$ and $|z_0\sin\theta| < 1.5~\mathrm{mm}$, where $d_0$ and $z_0$ are the transverse and longitudinal impact parameters.
  \item[(4)] There should be no other tracks with $\pt>0.5\GeV$ within a cone of radius $\Delta R=0.05$.
  \item[(5)] The candidate track should have the highest $\pt$ among the isolated tracks in the event and have $\pt$ above $10\GeV$.
  \item[(6)] The track should point to the TRT barrel layers and not point to the inactive regions around $\eta=0$.
  \item[(7)] The number of hits in the TRT outer module associated to the track ($\nTRT$) should be fewer than five.
\end{itemize}
The first four criteria were applied to all tracks in the event in order to ensure a well-reconstructed primary track whereas the fifth is meant to select chargino tracks that usually have the highest $\pt$ in the event.
The chargino tracks sufficiently fulfill the fifth criterion.
The sixth criterion was based on the extrapolated track position, and was set to avoid inactive regions of the TRT.
This requirement helped to reject fake disappearing tracks and works as an effective acceptance cut of $|\eta|<0.63$.
For the seventh criterion, $\nTRT$ was calculated by counting TRT hits lying on the extrapolated track.
The hits satisfying $d<r_{\mathrm{straw}}$ were taken into account, where $d$ is the distance between the hit and the track in the transverse plane and $r_{\mathrm{straw}}$ is the radius of the straw tube.
Hereafter, unless explicitly stated otherwise, "high-$\pt$ isolated track selection" and "disappearing track selection" indicate criteria (1)--(6) and (1)--(7), respectively.
Figure~\ref{figure:nTRT_data} shows the $\nTRT$ distributions with the high-$\pt$ isolated track selection requirements for data, signal and SM MC events.
When charginos decay before reaching the TRT outer module, $\nTRT$ is expected to have a value near zero; conversely, SM charged particles traversing the TRT typically have $\nTRT$ $\simeq$ 15.
The sample of selected tracks after requirements (1)--(6) is dominated by through-going tracks with $\nTRT$ $\simeq$ 15.
Criterion (7) removes the vast majority of these tracks: although it reduces the signal efficiency, it enhances the expected signal to background ratio very strongly.
These criteria select charginos decaying in the region $514<r<863~\mathrm{mm}$ effectively.
The data reduction is summarized in Table~\ref{table:data_reduction}.
After the application of all kinematic and track selection criteria, 185 candidate events remained. 

\begin{figure}
  \includegraphics[width=0.48\textwidth]{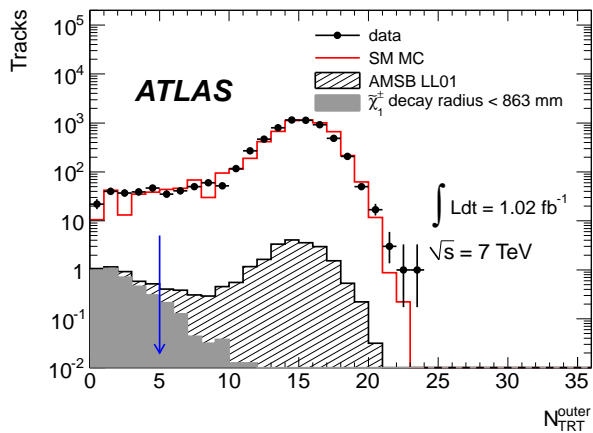}
  \caption{The $\nTRT$ distribution for data and signal events (LL01, $\tau_{\chinoonepm}=1~\mathrm{ns}$) with the high-$\pt$ isolated track selection. The selection boundary is indicated by the arrow.
  The expectation from SM MC events, normalized to the number of observed events, is also shown.
  When charginos decay before reaching the TRT outer module, $\nTRT$ is expected to have a value near zero; conversely, SM charged particles traversing the TRT typically have $\nTRT \simeq 15$.}
  \label{figure:nTRT_data}
\end{figure}

\begin{table}
\caption{Summary of selection cuts and data reduction. The selection efficiencies for each AMSB signal model are also shown.}
\label{table:data_reduction}
\begin{tabular}{m{32mm}<{\hfill}rrrr}
\hline
\hline
Selection & Data & \multicolumn{3}{c}{Signal efficiency~[\%]} \\
 & & LL01 & LL02 & LL03 \\
\hline
Trigger selection and non-collision rejection   & 1491012 & 90.2 & 90.2 & 89.3 \\
$e/\mu$ veto & 1390171 & 77.1 & 75.2 & 73.7 \\
$\met>130\GeV$ &   80971 & 67.9 & 68.8 & 69.4 \\
Jet $\pt$ requirements  &   18345 & 66.5 & 68.1 & 68.8 \\
High-$\pt$ isolated track & 6042 & 40.8 & 42.9 & 43.5 \\
Disappearing track & 185 & 6.8 & 7.5 & 7.4 \\
\hline
\hline
\end{tabular}
\end{table}

\section{Background estimation}
With the selection criteria described above, there are two main background sources for high-$\pt$ disappearing tracks:
\begin{itemize}
\item Charged hadrons (mostly charged pions) interacting with material in the TRT detector.
\item Low-$\pt$ charged particles whose $\pt$ is badly measured due to scattering in the inner detector material.
\end{itemize}
The two categories are labelled as ``high-$\pt$ interacting hadron track'' and ``bad track'' backgrounds, respectively. 
Fig.~\ref{figure:illustration_dissapearingtrack} shows schematically the origins of disappearing high-$\pt$ tracks.
According to the MC simulation, high-$\pt$ interacting hadron tracks were responsible for more than 95\% of the background tracks.
Electrons having low $\pt$ can be classified as disappearing tracks due to bremsstrahlung, however, the contribution of these tracks was negligibly small after the lepton veto and the track selection criterion (5).
\begin{figure}[htbp]
  \includegraphics[width=0.48\textwidth]{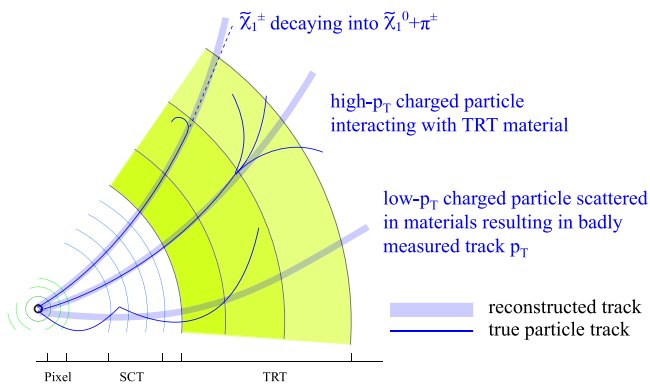}
  \caption{Origins of disappearing high-$\pt$ tracks.}
  \label{figure:illustration_dissapearingtrack}
\end{figure}

The fraction of events containing these background tracks is expected to be $\sim 10^{-4}$;
background estimation based on the MC simulation would therefore suffer from large uncertainties due to the lack of sufficient MC statistics and also from the difficulty in simulating the properties of these background mechanisms.
A data-driven background estimation technique was therefore used to estimate the background track $\pt$ spectrum, which used control samples enriched in the two background categories.
The main contribution to the high-$\pt$ interacting hadron background originated from charged hadrons in jets and $\tau$ hadronic decays.
In the $\pt$ range above $10\GeV$, where inelastic interactions dominate, the interaction rate has nearly no $\pt$-dependence \cite{Nakamura:2010zzi}.
Therefore, the $\pt$ spectrum of interacting hadron tracks was obtained from that of non-interacting hadron tracks.
By adopting the same kinematic selection criteria as those for the signal and ensuring penetration through the TRT detector by requiring $\nTRT>10$, a pure sample of high-$\pt$ non-interacting hadron tracks was obtained.
The contamination from bad tracks and any chargino signal was removed by requiring the calorimeter activity associated to the track, $\sum_{\Delta R<0.1}$ $\ET^{\mathrm{clus}}$ $/$ $ \pt^{\mathrm{track}}$, to be larger than 0.3, where $\pt^{\mathrm{track}}$ is the $\pt$ of the track and $\sum_{\Delta R<0.1}\ET^{\mathrm{clus}}$ is the sum of cluster transverse energies in a cone of $\Delta R=0.1$ around the track.
Simulation studies indicated that the $\pt$ spectrum of bad tracks depends little on the production process.
A sample with an enhanced bad track contribution was therefore obtained with the same track quality requirements as for the chargino track, but requiring $\met<100\GeV$.
The $\met$ requirement makes this sample orthogonal to the signal search sample.
In addition, the number of pixel hits associated to the track was required to be zero, and $\sum_{\Delta R<0.1} \ET^{\mathrm{clus}}/\pt^{\mathrm{track}}<0.3$ in order to reject possible contributions from high-$\pt$ interacting hadron tracks and to enhance the purity of bad tracks.
The requirement on the number of pixel hits had negligible impact on the shape of the reconstructed $\pt$ spectrum. 
The purity of bad tracks was close to 100\% after these requirements.

An ansatz functional form $(1+x)^{a_0}/x^{a_1+a_2\ln(x)}$ was fitted to the $\pt$ spectrum of the control sample of the high-$\pt$ non-interacting hadron tracks, where $x\equiv \pt^{\mathrm{track}}$ and $a_{i}$ $(i=0,1,2)$ are fit parameters.
Fig.~\ref{figure:backgroundtrack_fit}(a) shows the track $\pt$ distribution and the shape derived from a maximum likelihood fit.
Alternative fit functions gave shapes that agreed with each other and with the original form within the fit uncertainties.
The choice of functional form in this analysis was based on the $\chi^2$ values.

Bad tracks could have anomalously high values of $\pt$ and become a significant background.
Therefore, for the bad track background shape, a flat term representing the high-$\pt$ tail was added to give an estimate in the region of interest.
The resulting functional form was $(1+x)^{b_0}/x^{b_1+b_2\ln(x)}+b_3$, where $b_{i}$ $(i=0,1,2,3)$ are fit parameters.
The shape of the bad track background is shown in Fig.~\ref{figure:backgroundtrack_fit}(b).

\begin{figure*}
  \subfigure[High-$\pt$ hadron track sample]{
    \includegraphics[width=0.48\textwidth]{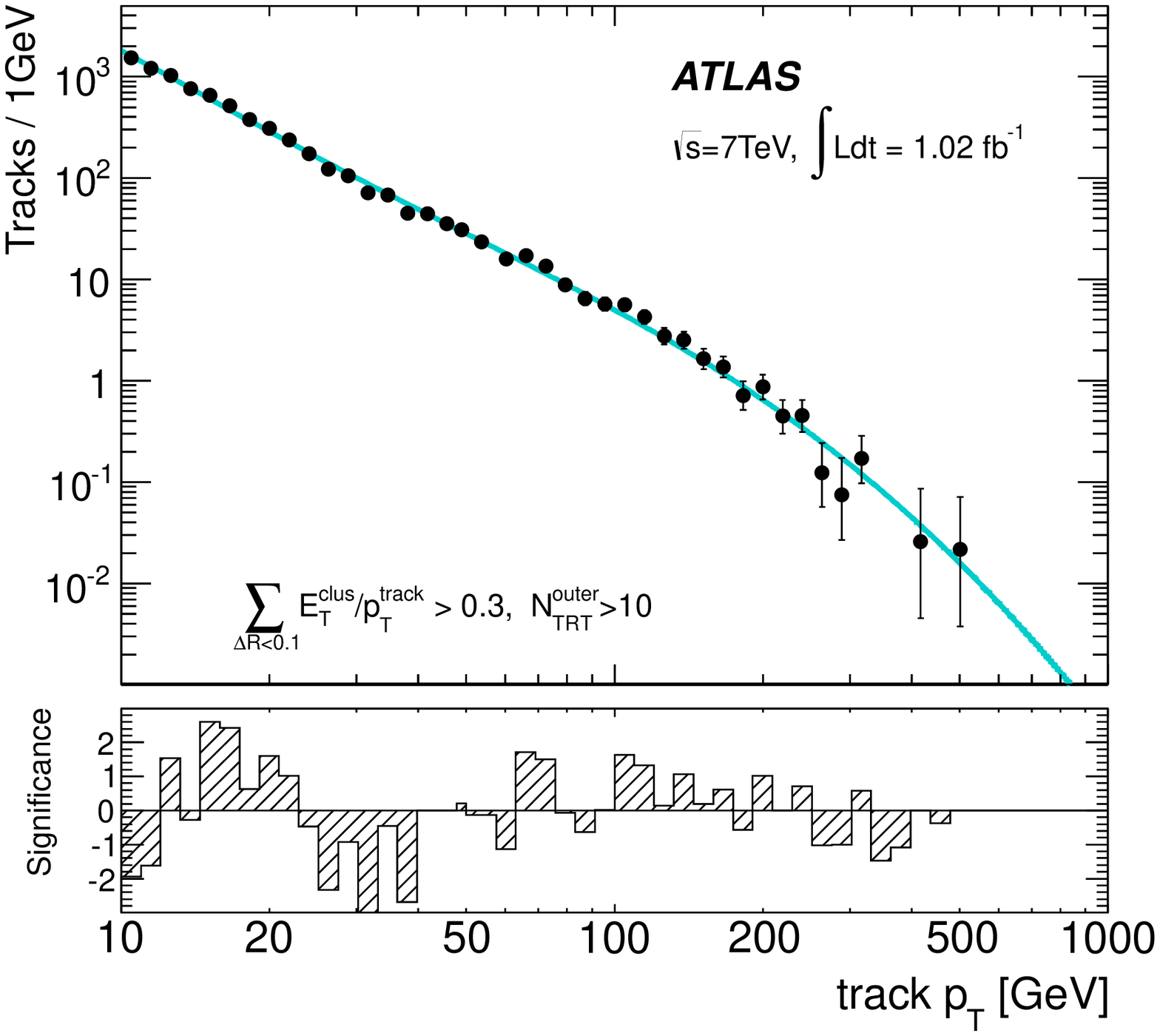}
  }
  \subfigure[Bad track sample]{
    \includegraphics[width=0.48\textwidth]{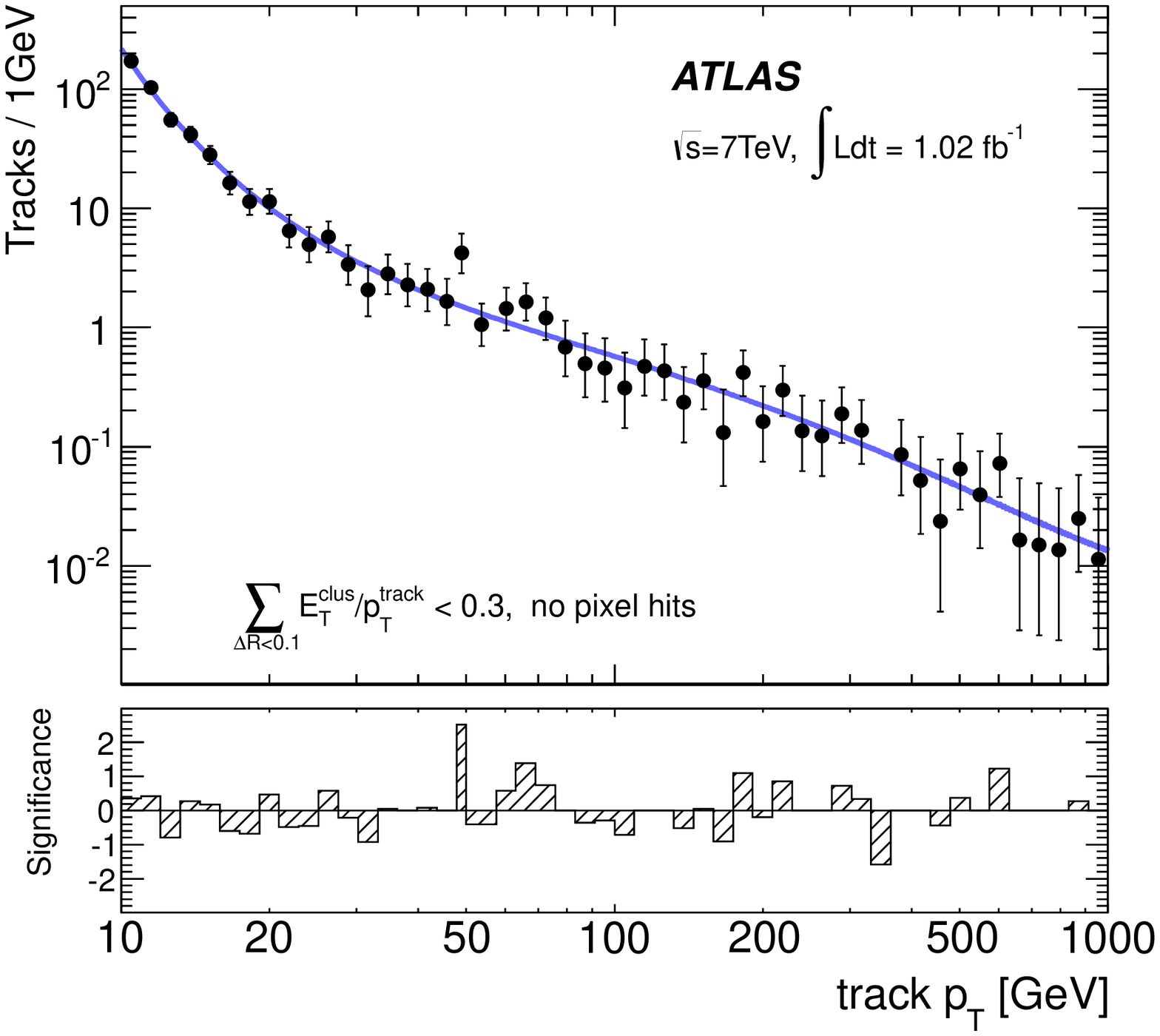}
  }
  \caption{The $\pt$ distributions of high-$\pt$ hadron track (a) and bad track (b) background control samples.
  The data and the fitted model are shown by the solid circles and the line, respectively.
  The significance of the data-model difference on a bin-by-bin basis is also shown at the bottom of each figure.} 
  \label{figure:backgroundtrack_fit}
\end{figure*}

\section{Signal extraction and constraints on the AMSB chargino}
In order to evaluate how well the observed data agree with a given signal model, a statistical test was performed based on a maximum likelihood.
The likelihood function for the sample of observed events ($n_{\mathrm{obs}}$), using the track $\pt$,  is defined as:
\begin{equation}
  \prod^{n_{\mathrm{obs}}} \frac{  \mu_{\mathrm{s}} n_{\mathrm{s}}^\mathrm{exp} \mathcal{L}_{\mathrm{s}} + n_{\mathrm{b}}\left\{ \left( 1-f_{\mathrm{bad}} \right) \mathcal{L}_{\mathrm{had}} + f_{\mathrm{bad}}\mathcal{L}_{\mathrm{bad}} \right\} }{n_{\mathrm{b}}+\mu_{\mathrm{s}} n_{\mathrm{s}}^\mathrm{exp}},
\end{equation}
where $\mu_{\mathrm{s}}$, $n_{\mathrm{s}}^{\mathrm{exp}}$, $n_{\mathrm{b}}$ and $f_{\mathrm{bad}}$ are
the signal strength (i.e. the ratio of a given cross section to its predicted value), 
the expected number of signal events for a given model,
the number of background events
and the fraction of bad tracks in the background, respectively.
The parameters $\mu_{\mathrm{s}}$, $n_{\mathrm{b}}$ and $f_{\mathrm{bad}}$ were left free in the fit.
The probability density functions of signal, interacting hadron track and bad track, $\mathcal{L}_{\mathrm{s}}$, $\mathcal{L}_{\mathrm{had}}$ and $\mathcal{L}_{\mathrm{bad}}$, are shown in Fig.~\ref{figure:pdfs}. 
The full shape of the distributions for $\pt>10\GeV$ was fitted with the two background contributions, and a signal contribution was also included in the fit for $\pt>50\GeV$.
A small signal contribution below $\pt=50\GeV$ was neglected.
The effects of systematic uncertainties were incorporated via constraint terms on nuisance parameters.
The overall normalisation of the signal and the parameters describing the background track $\pt$ shapes were set as nuisance parameters;
they were treated with a normal distribution and multivariate normal distributions with covariance matrices obtained by the fit of the background control samples, respectively.

\begin{figure}[htbp]
  \includegraphics[width=0.48\textwidth]{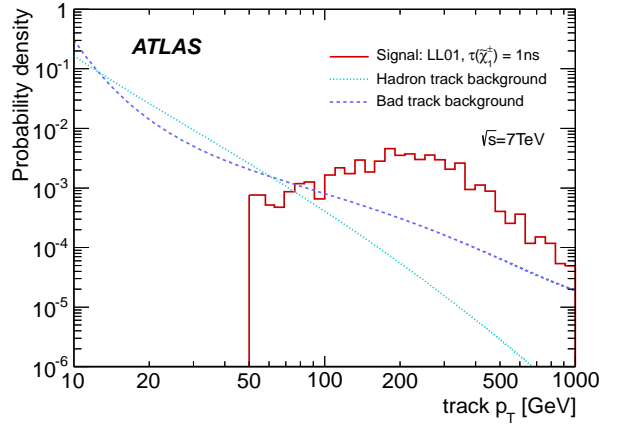}
  \caption{Probability densities of the signal (LL01, $\tau_{\chinoonepm}=1~\mathrm{ns}$) and background components, shown as a function of track $\pt$.
  In the signal case, only the region $\pt > 50 \GeV$ is shown.}
  \label{figure:pdfs}
\end{figure}

A total uncertainty of $\pm25\%$ was found for the signal normalisation;
the main contribution comes from the uncertainties in the theoretical cross section from the renormalisation and factorisation scales ($\pm 18\%$) and the parton distribution functions ($\pm 9\%$).
The jet energy scale~\cite{ATLAS-JET-ENERGY}, the track reconstruction efficiency~\cite{Aad:2011ir} and the integrated luminosity~\cite{Aad:2011dr,LuminosityGroup:1367408} could alter the signal yield; their contributions were estimated to be $\pm 9\%$, $\pm 2\%$ and $\pm 3.7\%$, respectively.
The systematic uncertainties due to pile-up were evaluated by examining the stability of the signal acceptance and the $\pt$ spectra of background tracks as a function of the number of $pp$ interactions.
Both data and signal MC were used for this purpose, and the resulting uncertainties were found to be negligible.

Fig.~\ref{figure:trackpt} shows the best-fit shape of the "signal + background" model for the sample signal point LL01 with $\tau_{\chinoonepm} = 1~\mathrm{ns}$ ($n_{\mathrm{s}}^{\mathrm{exp}}=4.2$).
The fit resulted in $n_{\mathrm{b}} = 185 \pm 14$ and the best fit values of $\mu_{\mathrm{s}}$ and $f_{\mathrm{bad}}$ were zero; upper limits of $\mu_{\mathrm{s}}<0.15$ and $f_{\mathrm{bad}}< 4.0 \times 10^{-2}$ were set at 68\% CL.
The $p$-value for the consistency of the observed data with the background-only hypothesis was calculated to be 0.5, showing that the observed track $\pt$ spectrum was in agreement with the background expectation.
The result also indicated that interacting hadron tracks were the dominant background, consistent with MC predictions.

\begin{figure}[htbp]
  \includegraphics[width=0.48\textwidth]{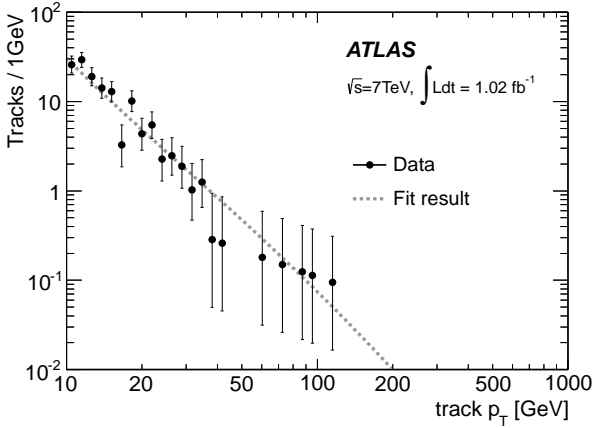}
  \caption{The $\pt$ distribution of candidate tracks with the best-fit shape of the "signal + background" model.
  The signal point of LL01 and $\tau_{\chinoonepm} = 1~\mathrm{ns}$ are used, but the best-fit signal contribution was found to be zero.}
  \label{figure:trackpt}
\end{figure}

The expected background and observed events in the region $\pt>50\GeV$ were $13 \pm 1$ and 5, respectively; this background estimate was derived from the background-only fit in the region $10<\pt\le 50\GeV$.
Model-independent upper limits were set on the cross section times acceptance for non-SM processes with the final state satisfying the kinematic and track selection criteria.
Fig.~\ref{figure:limit_xsecXacc} shows 95\% CL upper limits on the cross section times acceptance for candidate tracks with $\pt>\pt^0$ as a function of $\pt^0$.
The 95\% CL upper limit on the cross section for a given model was set by the point where the CL of the signal + background hypothesis based on the profile likelihood ratio \cite{Cowan:2010js} and the $CLs$ method \cite{Junk:1999kv,CLs} falls below 5\% when scanning the CL along various values of $\mu_{\mathrm{s}}$.
Fig.~\ref{figure:xsec_limit_LL01} shows the observed limit on the signal cross section at 95\% CL as a function of $\tau_{\chinoonepm}$ for the signal model LL01.
Limits on the chargino lifetime were also set: $\tau_{\chinoonepm} < 0.2$ or $\tau_{\chinoonepm} > 4~\mathrm{ns}$ for a chargino with a mass of $90\GeV$.
Moreover, a constraint on the chargino mass and lifetime was set by the scan of the observed cross section limits for the benchmark models, as shown in Fig.~\ref{figure:masslimit_life}.
In the framework of minimal AMSB with $m_{3/2}<32\TeV$, $m_0<1.5\TeV$, $\tan\beta=5$ and $\mu>0$, a chargino with $m_{\chinoonepm} < 92\GeV$ and $0.5 < \tau_{\chinoonepm} < 2~\mathrm{ns}$ was excluded at 95\% CL.

\begin{figure}[htbp]
  \includegraphics[width=0.48\textwidth]{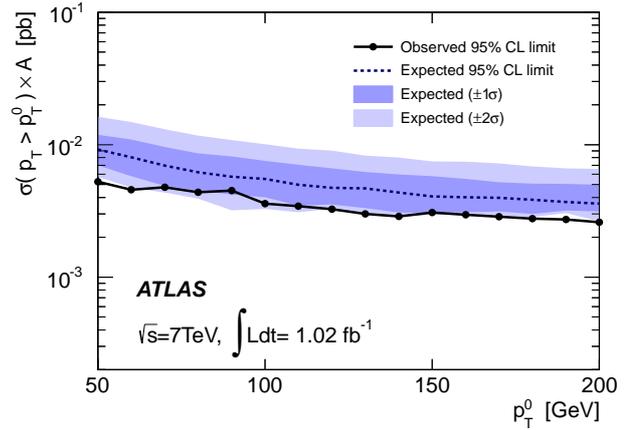}
  \caption{Model-independent upper limits on the cross section ($\sigma$) times acceptance ($A$) for a non-SM physics process containing an isolated, disappearing track with $\pt>\pt^0$ as a function of $\pt^0$.
  The observed and expected bounds at 95\% CL are shown. }
  \label{figure:limit_xsecXacc}
\end{figure}

\begin{figure}[htbp]
  \includegraphics[width=0.48\textwidth]{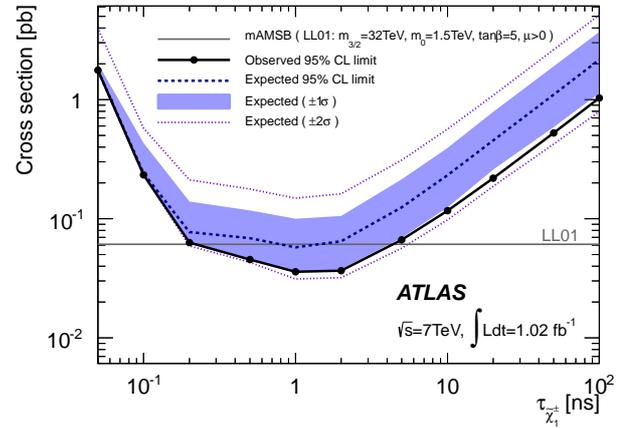}
  \caption{The observed and expected 95\% CL upper limits on the signal cross section as a function of chargino lifetime for $m_{\chinoonepm}=90.2\GeV$.
  The band and the dotted line indicate the range in which the limit is expected to lie due to the fluctuations in the expected background.}
  \label{figure:xsec_limit_LL01}
\end{figure}

\begin{figure}[htbp]
  \includegraphics[width=0.48\textwidth]{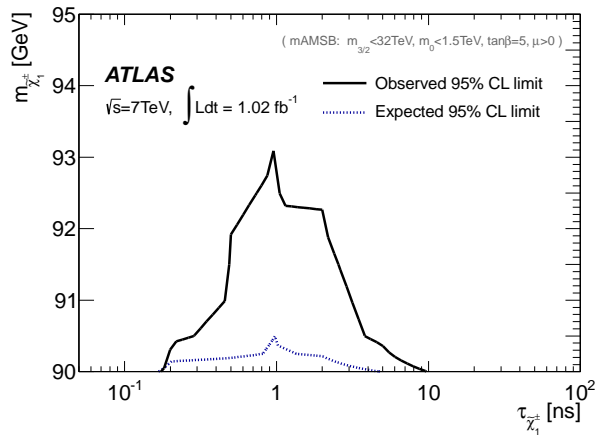}
  \caption{The constraint on the chargino mass and lifetime in a minimal AMSB model with $m_{3/2}<32\TeV$, $m_0 < 1.5\TeV$, $\tan\beta=5$ and $\mu>0$.
  The observed and expected bounds at 95\% CL are shown. }
  \label{figure:masslimit_life}
\end{figure}

\section{Conclusion}
The results of a search for long-lived charginos in $pp$ collisions with the ATLAS detector using $1.02~\ifb$ of data were presented in the context of AMSB scenarios.
The analysis used a signature of high-$\pt$ isolated tracks with few associated hits in the outer part of the ATLAS tracking system.
The $\pt$ spectrum of observed candidate tracks was found to be consistent with the expectation from SM background processes.
Constraints on the AMSB chargino mass and lifetime were set; a chargino having $m_{\chinoonepm} < 92\GeV$ and $0.5 < \tau_{\chinoonepm} < 2~\mathrm{ns}$ was excluded at 95\% CL.

\begin{acknowledgements}




We thank CERN for the very successful operation of the LHC, as well as the
support staff from our institutions without whom ATLAS could not be
operated efficiently.

We acknowledge the support of ANPCyT, Argentina; YerPhI, Armenia; ARC,
Australia; BMWF, Austria; ANAS, Azerbaijan; SSTC, Belarus; CNPq and FAPESP,
Brazil; NSERC, NRC and CFI, Canada; CERN; CONICYT, Chile; CAS, MOST and NSFC,
China; COLCIENCIAS, Colombia; MSMT CR, MPO CR and VSC CR, Czech Republic;
DNRF, DNSRC and Lundbeck Foundation, Denmark; EPLANET and ERC, European Union;
IN2P3-CNRS, CEA-DSM/IRFU, France; GNAS, Georgia; BMBF, DFG, HGF, MPG and AvH
Foundation, Germany; GSRT, Greece; ISF, MINERVA, GIF, DIP and Benoziyo Center,
Israel; INFN, Italy; MEXT and JSPS, Japan; CNRST, Morocco; FOM and NWO,
Netherlands; RCN, Norway; MNiSW, Poland; GRICES and FCT, Portugal; MERYS
(MECTS), Romania; MES of Russia and ROSATOM, Russian Federation; JINR; MSTD,
Serbia; MSSR, Slovakia; ARRS and MVZT, Slovenia; DST/NRF, South Africa;
MICINN, Spain; SRC and Wallenberg Foundation, Sweden; SER, SNSF and Cantons of
Bern and Geneva, Switzerland; NSC, Taiwan; TAEK, Turkey; STFC, the Royal
Society and Leverhulme Trust, United Kingdom; DOE and NSF, United States of
America.

The crucial computing support from all WLCG partners is acknowledged
gratefully, in particular from CERN and the ATLAS Tier-1 facilities at
TRIUMF (Canada), NDGF (Denmark, Norway, Sweden), CC-IN2P3 (France),
KIT/GridKA (Germany), INFN-CNAF (Italy), NL-T1 (Netherlands), PIC (Spain),
ASGC (Taiwan), RAL (UK) and BNL (USA) and in the Tier-2 facilities
worldwide.


\end{acknowledgements}

\clearpage 
\onecolumn
\begin{flushleft}
{\Large The ATLAS Collaboration}

\bigskip

G.~Aad$^{\rm 48}$,
B.~Abbott$^{\rm 110}$,
J.~Abdallah$^{\rm 11}$,
A.A.~Abdelalim$^{\rm 49}$,
A.~Abdesselam$^{\rm 117}$,
O.~Abdinov$^{\rm 10}$,
B.~Abi$^{\rm 111}$,
M.~Abolins$^{\rm 87}$,
H.~Abramowicz$^{\rm 152}$,
H.~Abreu$^{\rm 114}$
% ...
\bigskip

% ...
$^{10}$ Institute of Physics, Azerbaijan Academy of Sciences, Baku, Azerbaijan\\
$^{11}$ Institut de F\'isica d'Altes Energies and Departament de F\'isica de la Universitat Aut\`onoma  de Barcelona and ICREA, Barcelona, Spain\\
% ...
$^{48}$ Fakult\"{a}t f\"{u}r Mathematik und Physik, Albert-Ludwigs-Universit\"{a}t, Freiburg i.Br., Germany\\
$^{49}$ Section de Physique, Universit\'e de Gen\`eve, Geneva, Switzerland\\
% ...
$^{87}$ Department of Physics and Astronomy, Michigan State University, East Lansing MI, United States of America\\
% ...
$^{110}$ Homer L. Dodge Department of Physics and Astronomy, University of Oklahoma, Norman OK, United States of America\\
$^{111}$ Department of Physics, Oklahoma State University, Stillwater OK, United States of America\\
% ...
$^{114}$ LAL, Univ. Paris-Sud and CNRS/IN2P3, Orsay, France\\
% ...
$^{117}$ Department of Physics, Oxford University, Oxford, United Kingdom\\
% ...
$^{152}$ Raymond and Beverly Sackler School of Physics and Astronomy, Tel Aviv University, Tel Aviv, Israel\\
% ...
\end{flushleft}


\end{document}